\documentstyle[12pt]{article}
\begin{document}

\begin{center}
\vspace*{1cm}{\Large {\bf \ Theoretical investigation on the possibility of
preparing left-handed materials in metallic magnetic granular composites }}
\medskip \\[0pt]
\smallskip S.T.Chui and Liangbin Hu \\[0pt]
\smallskip

\vspace*{0.2cm} {\it Bartol Research Institute, University of Delaware,
Newark, Delaware, U.S.A} \vspace*{0.2cm}
\end{center}

\centerline{\bf\large Abstract} \vspace*{1cm} We
investigate the possibility of preparing left-handed materials in metallic
magnetic granular composites. \ Based on the effective medium 
approximation, \ we show
that by incorporating metallic magnetic nanoparticles into an appropriate
insulating matrix and controlling the directions of magnetization of
metallic magnetic components and their volume fraction, \ it may be possible
to prepare a composite medium of low eddy current loss which is
left-handed for electromagnetic waves propagating in some special direction
and polarization in a  frequency region near the ferromagnetic resonance
frequency.
This composite may be easier to make on an industrial scale. In addition, its
physical properties may be easily tuned by rotating the magnetization locally.
\medskip

 \vspace*{1cm}

\newpage

In classical electrodynamics, the response (typically frequency-dependent) of
a material to electric and magnetic fields is characterized by two
fundamental quantities, \ the permittivity $\epsilon $ and the permeability $%
\mu $. \ The permittivity relates the electric displacement field $\vec{D}$
to the electric field $\vec{E}$ through $\vec{D}=\epsilon \vec{E}$, \ and
the permeability $\mu $ relates the magnetic field $\vec{B}$ and $\vec{H}$ \
by $\ \vec{B}=\mu \vec{H}$. \ If we do not take losses into account and
treat $\epsilon $ and $\mu $ as real numbers, \ according to Maxwell$%
^{\prime }$s equations, \ electromagnetic waves can propagate through a
material only if the index of refraction $n$, \ given by $(\epsilon \mu
)^{1/2}$, \ is real. \ \ (Dissipation will add imaginary components to $%
\epsilon $ and $\mu $ and cause losses, \ but for a qualitative picture, \
one can ignore losses and treat $\epsilon $ and $\mu $ as real numbers. \
Also, strictly speaking, $\ \epsilon $ and $\mu $ are second-rank tensors, \
but they reduce to scalars for isotropic materials). \ In a medium with$%
\epsilon $ and $\mu $ both positive, the index of refraction is real and
electromagnetic waves can propagate. \ All our everyday transparent
materials are such kind of media. \ In a medium where one of $\epsilon $ and $%
\mu $ is negative but the other is positive, \ the index of refraction is
imaginary and electromagnetic waves cannot propagate. \ Metals and Earth$%
^{\prime }$s ionosphere are such kind of media. Metals and the ionosphere
have free electrons that have a natural frequency-the plasma frequency-which
is on the order of 10MHz in the ionosphere and falls at or above visible
frequencies for most metals. \ At frequencies above the plasma frequency, $\
\epsilon $ is positive and electromagnetic waves are transmitted. \ For
lower frequencies, $\ \epsilon $ becomes negative and the index of
refraction is imaginary and consequently electromagnetic waves cannot
propagate through. \ In fact, the electromagnetic response of metals in the
visible and near ultraviolet regions is dominated by the negative epsilon
concept[1]-[4].

Although all our everyday transparent materials have both positive $\epsilon 
$ and positive $\mu $, \ from the 
theoretical point of view, \ in a medium with $%
\epsilon $ and $\mu $ both negative, \ the index of refraction is also
positive and electromagnetic waves can also propagate through, moreover, \
if such media exist, \ the propagation of waves through them should give
rise to several peculiar properties. \ This was first pointed out by 
Veselago over 30 thirty years ago-when no material with simultaneously
negative $\epsilon $ and $\mu $ was known[5]. \ For example, \ the cross
product of $\vec{E}$ and $\vec{H}$ for a plane wave in regular media gives
the direction of both propagation and energy flow, \ and the electric field $%
\vec{E}$, \ the magnetic field $\vec{H}$, \ and the wave vector 
$\vec{k}$ form a
right-handed triplet of vectors. \ In contrast, \ in a medium with $\epsilon 
$ and $\mu $ both negative, \ $\vec{E}\times \vec{H}$ for a plane wave still
gives the direction of energy flow, \ but the wave itself (that is, the phase
velocity) propagates in the opposite direction, \ i.e., \ wave vector $\vec{k%
}$ lies in the opposite direction of $\vec{E}\times \vec{H}$ for propagating
waves. \ In this case, \ electric field $\vec{E}$, \ magnetic field $\vec{H}$%
, \ and wave vector $\vec{k}$ \ form a left-handed triplet of vectors.
Such a medium is therefore termed left-handed medium[5]. \ In addition to
this $^{^{\prime }}$left-handed$^{\prime }$ characteristic, \ there are a
number of other dramatically different propagation characteristics stemming
from a simultaneous change of the signs of $\epsilon $ and $\mu $, \
including reversal of both the Doppler shift and the Cerenkov radiation, \ anomalous
refraction, \ and even reversal of radiation pressure to radiation tension.
\ However, \ although these counterintuitive properties follow directly from
Maxwell$^{\prime }$s equations-which still hold in these unusual materials,
\ such type of left-handed materials have never been found in nature and
these peculiar propagation properties have never been demonstrated
experimentally. \ If such media can be prepared artificially, \ they will
offer exciting opportunities to explore new physics and potential
applications in the area of radiation-material interactions. \ Recently, \
interesting progress has been achieved in preparing a $^{^{\prime }}$%
left-handed$^{^{\prime }}$ material artificially. \ 
Following the suggestion of Pendry[1],
Smith and coworkers reported that a medium made up of an array of
conducting nonmagnetic split ring resonators and continuous thin wires can
have both an effective negative permittivity $\epsilon $ and negative
permeability $\mu $ \ for electromagnetic waves propagating in some special
direction and special polarization at microwave frequencies[6]. \ This is
the first experimental realization of an artificial preparation
of a left-handed material. \ Motivated by this progress, \ in this rapid communication, \
we propose to investigate the possibility of preparing left-handed materials
in another type of systems-metallic magnetic granular composites. \ The idea
is that, \ by incorporating metallic ferromagnetic nanoparticles into an
appropriate insulating matrix and controlling the directions of
magnetization of \ metallic magnetic particles and their volume fraction, \
it may be possible to achieve a composite medium that has simultaneously
negative $\epsilon $ and negative $\mu $ and low eddy current loss. \ This
idea was based on the fact that on the one hand, \ the permittivity of metallic
particles is automatically negative at frequencies less than the plasma
frequency, \ and on the other hand, \ the effective permeability of \
ferromagnetic materials for electromagnetic waves propagating in some
particular direction and polarization can be negative at frequency in the
vicinity of the ferromagnetic resonance frequency $\omega _{0},$ which is
usually in the frequency region of microwaves. \ So, \ if we can prepare a
composite medium in which one component is both metallic and ferromagnetic
and other component insulating, \ and we can control the directions of
magnetization of \ metallic magnetic particles and their volume fraction, \
it may be possible to achieve a left-handed composite medium of low eddy
current losses for electromagnetic waves propagating in some special
direction and polarization. \ 
This composite may be easier to make on an industrial scale. In addition, its
physical properties may be easily tuned by rotating the magnetization locally.

To illustrate the above idea more clearly, \ in the following we present 
results of calculations 
based on the effective medium theory. \ Let us consider an
idealized metallic magnetic granular composite consisting of two types of
spherical particles, \ in which \ one type of particles are metallic
ferromagnetic grains of radius $R_{1}$, \ the other type are non-magnetic
dielectric (insulating) grains of radius $R_{2}$. \ Each grain is assumed to
be homogeneous. \ The directions of magnetization of all metallic magnetic
grains are assumed to be in the same direction. In 
length scales much larger than the grain sizes, \ the composite can be
considered as a homogeneous magnetic system. \ The permittivity and
permeability of non-magnetic dielectric grains are both scalars, and will be
denoted as $\epsilon _{1}$ and $\mu _{1}$.\ \ The permittivity of metallic
magnetic grains will be denoted as $\epsilon _{2}$ and will be taken to have
a Drude form: $\epsilon _{2}=1-\omega _{p}^{2}/\omega (\omega +i/\tau )$, \
\ where $\omega _{p}$ is the plasma frequency of the metal and $\tau $ is a
relaxation time. \ Such a form of $\epsilon $ is representative of a variety
of metal composites[8]-[9]. \ The permeability of metallic magnetic grains
are second-rank tensors and will be denoted as $\hat{\mu}_{2},$ \ which can
be derived from the Landau-Lifschitz equations[7]. \ Assuming that the
directions of magnetization of all magnetic grains are in the direction of \
the $z$-axis, \ $\hat{\mu}_{2}$ will have the following form[7]: 
\begin{equation}
\hat{\mu _{2}}=\left[ 
\begin{array}{cccc}
\mu _{a} & -i\mu ^{\prime } & 0 &  \\ 
i\mu ^{\prime } & \mu _{a} & 0 &  \\ 
0 & 0 & 1 & 
\end{array}
\right]
\end{equation}
where 
\begin{eqnarray}
\mu _{a} &=&1+\frac{\omega _{m}(\omega _{0}+i\alpha \omega )}{(\omega
_{0}+i\alpha \omega )^{2}-\omega ^{2}}, \\
\mu ^{\prime } &=&-\frac{\omega _{m}\omega }{(\omega _{0}+i\alpha \omega
)^{2}-\omega ^{2}},
\end{eqnarray}
$\omega _{0}=\gamma \vec{H_{0}}$ is the ferromagnetic resonance
frequency, $\ H_{0}$ is the effective magnetic field in magnetic particles
and may be a sum of the external magnetic field, \ the effective anisotropy
field and the demagnetization field; $\ \omega _{m}=\gamma \vec{M_{0}}$, $\ $%
where $\gamma $ is the gyromagnetic ratio, $\ M_{0}$ is the saturation
magnetization of magnetic particles; \ $\alpha $ is the magnetic damping
coefficient; $\ \omega $\ is the frequency of incident electromagnetic
waves. \ We shall only consider incident electromagnetic waves propagating in
the direction of the magnetization. \ This is the most interesting case in the
study of magneto-optical effects in ferromagnetic materials. \ We also
assume that the grain sizes are much smaller compared with the characteristic
wavelength $\lambda ,$ and consequently, electromagnetic waves in the
composite can be treated as propagating in a homogeneous magnetic system. \
According to Maxwell$^{^{\prime }}$s equations, \ electromagnetic waves
propagating in the direction of magnetization in a homogeneous magnetic
material is either positive or negative transverse circularly polarized. \
If \ the composite can truly be treated as a homogeneous magnetic system in
the case of grain sizes much smaller than the characteristic wavelength, \
electric and magnetic fields in the composite should also be either positive
or negative circularly polarized and can be expressed as : 
\begin{eqnarray}
\vec{E}(\vec{r},t) &=&\vec{E}_{0}^{(\pm )}e^{ikz-\beta z-i\omega t} \\
\vec{H}(\vec{r},t) &=&\vec{H}_{0}^{(\pm )}e^{ikz-\beta z-i\omega t}
\end{eqnarray}
where $\vec{E}_{0}^{(\pm )}=\hat{x}\mp i\hat{y}$, $\ \vec{H}_{0}^{(\pm )}=%
\hat{x}\mp i\hat{y}$, $\ k={\rm Real}[k_{eff}]$ is the effective wave
number, $\ \beta ={\rm Im}[k_{eff}]$ is the effective damping coefficient
caused by the eddy current, $\ k_{eff}=k+i\beta $ is the effective propagation
constant. $\ $In Eqs.(4)-(5) the signs of $k$ and $\beta $ can both be
positive or negative depending on the directions of the wave vector and the 
energy
flow. $\ $For convenience we assume that the direction of energy flow is in
the positive direction of the $z$ axis, \ i.e., \ we assume $\beta >0$ in
Eqs.(4)-(5)$,$ but the sign of $k$ still can be positive or negative. \ In
this case, \ if $k>0$, \ the phase velocity and energy \ flow are in the
same directions, \ and \ from Maxwell$^{^{\prime }}$s equation, \ one can
see that the electric and magnetic field $\vec{E}$ and $\vec{H}$ and the
wave vector 
$\vec{k}$ will form a right-handed triplet of vectors. \ This is the usual
case for right-handed materials. In contrast, \ if $k<0,$ the phase velocity
and energy \ flow are in opposite directions, \ and $\vec{E}$, $\ \vec{H}
$ and $\vec{k}$ will form a left-handed triplet of vectors. \ This is just
the peculiar case for left-handed materials. \ So, \ for incident waves of a
given frequency $\omega ,$ we can determine whether wave propagations in the
composite is right-handed or left-handed through the relative sign changes
of $k$ and $\beta $. \ In the following, \ we shall determine the effective
propagation constant $k_{eff}=k+i\beta $ by means of the effective
medium approximation. \ The details of \ various kinds of effective medium
approximations have been discussed in a series of references[8]-[12], \ here
we only list the main points. \ First, \ if the composite can
truly be considered as a homogeneous magnetic system in the case of grain
sizes much smaller than the characteristic wavelength, \ then for waves
(positive or negative circularly polarized) propagating through the composite
in the direction of \ magnetization, \ \ their propagations can be described
by an effective permittivity $\epsilon _{eff}$ and an effective permeability 
$\mu _{eff}$, \ which satisfy the following relations 
\begin{eqnarray}
\int \vec{D}(\vec{r},\omega )e^{ik_{eff}z}d\vec{r} &=&\epsilon _{eff}\int 
\vec{E}(\vec{r},\omega )e^{ik_{eff}z}d\vec{r}, \\
\int \vec{B}(\vec{r},\omega )e^{ik_{eff}z}d\vec{r} &=&\mu _{eff}\int \vec{h}(%
\vec{r},\omega )e^{ik_{eff}z}d\vec{r},
\end{eqnarray}
where $k_{eff}$ and $\omega $ are related by $k_{eff}=\omega \lbrack
\epsilon _{eff}\mu _{eff}]^{1/2}$. \ \ Although these relations are simple
and in principle exact, \ it is very difficult to calculate the integrals in
them because the fields in the composite are spatially varying in a random
way. \ \ One therefore must resort to various types of approximations. \ The
simplest approximation is the effective medium approximation. \ In this
approximation, \ we calculate the fields in each grain as if the grain were
embedded in an effective medium of dielectric constant $\epsilon _{eff}$ \
and magnetic permeability $\mu _{eff}$. \ Consider, \ for example, \ the $i$%
th grain. Under the embedding assumption, \ the electric and magnetic fields
incident on the grain are the form of Eqs.(4)-(5): 
\begin{eqnarray}
\vec{E}_{inc} &=&\vec{E}_{0}^{(\pm )}e^{ik_{eff}z-i\omega t}, \\
\vec{h}_{inc} &=&\vec{h}_{0}^{(\pm )}e^{ik_{eff}z-i\omega t},
\end{eqnarray}
where $\vec{E}_{0}^{(\pm )}=\hat{x}\mp i\hat{y}$ and $\vec{h}_{0}^{(\pm )}=%
\hat{x}\mp i\hat{y},$ \ corresponding to the positive($+$) or negative($-$)
circularly polarized waves. \ If the fields inside the grain can be found, \
then the inside fields can be used to calculate the integral over the grain
volume 
\begin{eqnarray}
\vec{I}_{i} &=&\int_{v_{i}}\vec{E}_{i}(\vec{r},\omega )e^{ik_{eff}z}d\vec{r},
\\
\vec{J}_{i} &=&\int_{v_{i}}\vec{h}_{i}(\vec{r},\omega )e^{ik_{eff}z}d\vec{r},
\end{eqnarray}
which is required to find the integral in Eqs.(6)-(7). \ For the positive or
negative circularly polarized incident waves described by Eqs.(8)-(9), \ the
integral $\vec{I}_{i}$ and $\vec{J}_{i}$ can be written as 
\begin{eqnarray}
\vec{I}_{i} &=&(\hat{x}\mp i\hat{y})I_{i}, \\
\vec{J}_{i} &=&(\hat{x}\mp i\hat{y})J_{i},
\end{eqnarray}
where $I_{i}$ and $J_{i}$ are scalars. \ If $I_{i}$ and $J_{i}$ can be
found, \ then from Eqs.(6)-(7), \ the effective permittivity $\epsilon
_{eff} $ \ and effective permeability $\mu _{eff}$ \ can be calculated by 
\begin{equation}
\epsilon _{eff}=\frac{f_{1}\epsilon _{1}I_{1}+f_{2}\epsilon _{2}I_{2}}{%
f_{1}I_{1}+f_{2}I_{2}},
\end{equation}
\begin{equation}
\mu _{eff}=\frac{f_{1}\mu _{1}J_{1}+f_{2}\mu _{2}^{(\pm )}J_{2}}{%
f_{1}J_{1}+f_{2}J_{2}},
\end{equation}
where $f_{1}$ and $f_{2}$ are the volume fractions of the two types of
grains, $\ \mu _{1}$ is the permeability of \ non-magnetic dielectric
grains, $\mu _{2}^{(+)}=\mu _{a}-\mu ^{\prime }$ and $\mu ^{(-)}=\mu
_{a}+\mu ^{\prime }$ (see Eqs.1-3) are the effective permeability of
magnetic grains for positive and negative circularly polarized waves
respectively. \ As for the calculation of $I_{i}$ and $J_{i},$ \ we can
follow the method of expanding interior and exterior fields in a multipole
series and matching the boundary conditions[13]. \ After the coefficients of
the multipole expansion of interior and exterior fields are obtained by
matching the boundary conditions, \ $I_{i}$ and $J_{i}$ can be found
and subsequently be substituted into Eqs.(14)-(15). \ Since this method is
standard, \ we shall not present the details.\ \ In the final results, 
Eqs.(14)-(15) reduce to one self-consistent equation: 
\begin{eqnarray}
&&\sum_{i=1,2}f_{i}\sum_{l=1}^{\infty }(2l+1)[\frac{k_{eff}\psi _{l}^{\prime
}(k_{i}R_{i})\psi _{l}(k_{eff}R_{i})-k_{i}\psi _{l}(k_{i}R_{i})\psi
_{l}^{\prime }(k_{eff}R_{i})}{k_{eff}\psi _{l}^{\prime }(k_{i}R_{i})\zeta
_{l}(k_{eff}R_{i})-k_{i}\psi _{l}(k_{i}R_{i})\zeta _{l}^{\prime
}(k_{eff}R_{i})}  \nonumber \\
&&+\frac{k_{i}\psi _{l}^{\prime }(k_{i}R_{i})\psi
_{l}(k_{eff}R_{i})-k_{eff}\psi _{l}(k_{i}R_{i})\psi _{l}^{\prime
}(k_{eff}R_{i})}{k_{i}\psi _{l}^{\prime }(k_{i}R_{i})\zeta
_{l}(k_{eff}R_{i})-k_{eff}\psi _{l}(k_{i}R_{i})\zeta _{l}^{\prime
}(k_{eff}R_{i})}]=0,
\end{eqnarray}
where $R_{i}$ is the radius of the $i$th type of grains, \ and 
\begin{eqnarray}
k_{1} &=&\omega \lbrack \epsilon _{1}\mu _{1}]^{1/2}, \\
k_{2} &=&\omega \lbrack \epsilon _{2}\mu _{2}^{(\pm )}]^{1/2}, \\
\psi _{l}(x) &=&xj_{l}(x), \\
\zeta _{l}(x) &=&xh_{l}^{(1)}(x),
\end{eqnarray}
$j_{l}(x)$ and $h_{l}(x)$ are the usual spherical Bessel and Hankel
functions. \ Eq.(16) determine the effective product of $(\epsilon \mu
)_{eff},$ or equivalently $k_{eff},$ but not a single $\epsilon _{eff}$ and $%
\mu _{eff}.$ \ \ It can describe the change of the phase of a plane wave
across a slab of the composite, \ but it does not precisely describe wave
propagations across a slab of the composite. \ This is due to the fact we
make no attempt to rigorously solve the boundary-value problem for a slab of
composite by matching the fields inside the slab and external fields outside
the slab at the boundary. \ In fact, \ it is common in various types of
effective medium theories that for $\omega \neq 0$ the electromagnetic
properties of a composite cannot in general be specified by a single $%
\epsilon _{eff}$ and $\mu _{eff}.$ \ Since we can determine whether wave
propagations through the composite is left-handed or right-handed by the
calculation of the effective propagation constant $k_{eff},$ \ Eq.(16) is
enough for the problems we are discussing.

The numerical results for a metal volume fraction $f_2$ of 0.3
obtained from Eq.(16) are summarized in Fig.1-Fig.2. \
Fig.1(a) shows the frequency dependence of the real part of the effective
permeability $\mu ^{(+)}$ of magnetic grains for positive circularly
polarized plane waves, \ Fig.1(b) and (c) show the corresponding frequency
dependences of the effective wave number $k$ and the effective damping
coefficient $\beta $ in a composite consisting of metallic magnetic grains
and dielectric grains. \ From Eqs.(1)-(3), \ we can see that if the magnetic
damping coefficient $\alpha $ is zero, $\ {\rm Re}[\mu ^{(+)}]$ will be
negative in the whole frequency region of $\omega >\omega _{0}($the magnetic
resonance frequency). \ From Fig.1(a), \ we can see that if $\alpha $ is
nonzero but small enough, \ there can still be a  frequency region
near $\omega _{0}$ in which ${\rm Re}[\mu ^{(+)}]$ is negative. \ In this
case, \ if the amplitude of the negative $\mu ^{(+)}$ is large enough, \ \ $%
k $ will be negative in this  frequency region \ as was shown in
Fig.1(b), and hence the phase velocity and energy \ flow will be in the
opposite directions in this  frequency region, \ and $\vec{E}$, $\ 
\vec{H}$ and $\vec{k}$ will form a left-handed triplet of vectors, \ i.e.,
the composite will be left-handed in this  frequency region for
positive circularly polarized plane waves. \ But if $\alpha $ is not small
enough, $\ {\rm Re}[\mu ^{(+)}]$ will be positive in the whole frequency
region, \ or though $Re[\mu ^{(+)}]$ is negative in a  frequency
region near $\omega _{0}$, \ the amplitude of the negative $Re[\mu ^{(+)}]$
is not large enough, \ in this case $k$ will be positive in the whole
frequency region, \ as was shown in Fig.1(b). \ In this case, \ the
composite is right-handed for\ positive circularly polarized waves in the
whole frequency region. \ The calculations also show that if the radius of
metallic grains are small enough and the volume fraction of metal components
is smaller than the threshold value of the insulator-metal transition, \
which is approximately $1/3$ \ in our model, \ the losses caused by eddy
current are very small and the composite is essentially an insulator. This
can be seen from Fig.1(c), \ in which the damping coefficient $\beta $ is
very small compared with the amplitude of the wave number $k,$ \ i.e., \ the
eddy current losses are very small \ in the cases shown in Fig.1. If the
volume fraction of metal components is larger than the threshold value, \
the composite will be essentially a metal, \ and the damping coefficient $%
\beta $ will be much larger than the amplitude of wave number $k$(not shown
in the figure). \ In Fig.2(a) we show the frequency dependence of the real
part of the effective permeability $\mu ^{(-)}$ of magnetic grains for
negative circularly polarized waves, \ and in Fig.2(b) we show the
corresponding frequency dependence of the effective wave number $k$ in a
composite consisting of the metallic magnetic grains and dielectric grains.
We can see that for negative circularly polarized waves, $\ {\rm Re}[\mu
^{(-)}]$ is positive in the whole frequency region no matter how small $%
\alpha $ is, \ and correspondingly, \ $k$ is positive in the whole frequency
region, \ i.e., the composite is right-handed in the whole frequency region
for negative circularly polarized waves no matter how small $\alpha $ is.

In conclusion, \ we have discussed the possibility of preparing a
left-handed material in metallic magnetic granular composites based on the
effective medium approximation. 
Our model analysis shows that, \ by incorporating
metallic magnetic nanoparticles into an appropriated insulating matrix and
controlling the directions of magnetization of metallic magnetic components
and their volume fraction, \ it may be possible to prepare a composite
medium of low eddy current losses which is left-handed for electromagnetic
waves propagating in some special direction and polarization in a 
frequency region near the magnetic resonance frequency.
%\ Though this simple
%model analysis shows that there exists the possibility of preparing a
%left-handed medium in metallic magnetic granular composites, \ from the
%experimental point of view, \ there are a lot of doubts on this possibility
%because the conditions assumed in the theoretical calculations may be rather
%difficult to be satisfied in real experiments. \ Of course, \ the validity
%of the simple effective medium approximation applied in this paper may also
%be a question.
\ Further theoretical investigations by
%improved effective medium type of \ theories and
other approximations such as first-principle numerical calculations may be
needed to further confirm the possibility shown in this paper.

S. T. Chui was supported in part by the Office of Naval Research and by the
Army Research Laboratory through the Center of Composite Materials at the
University of Delaware. We thank John Xiao for helpful discussions.

{\Large \newpage }

\begin{center}
{\normalsize Caption }
\end{center}

{\normalsize \bigskip }

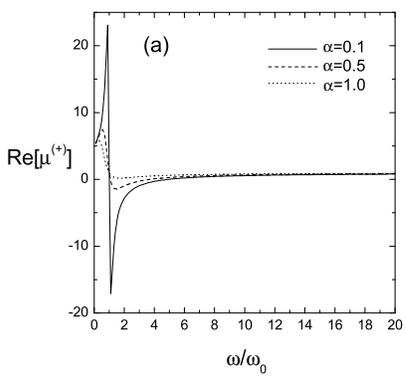
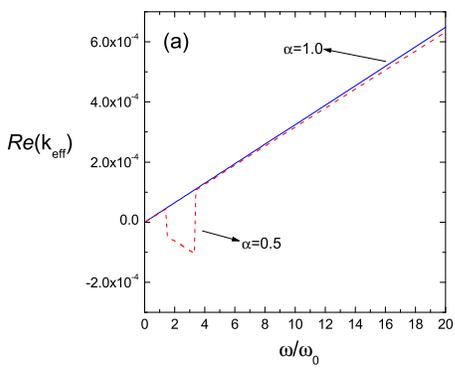
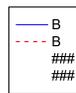
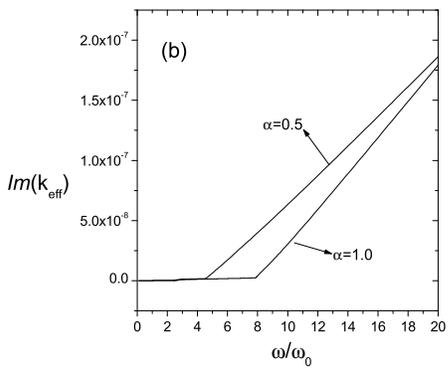
{\normalsize \noindent Fig.1. (a) The frequency dependencies of the
effective permeability }$\mu ^{(+)}${\normalsize \ of magnetic grains and
the corresponding frequency dependencies of (b) the effective wave number }$%
k ${\normalsize \ and (c) the effective damping coefficient }$\beta $%
{\normalsize \ of the composite for positive circularly polarized waves
propagating in the direction of magnetization. \ (The plasma frequency }$%
\omega _{p}${\normalsize \ is usually in the visible or ultraviolet
frequency region and the ferromagnetic resonance frequency }$\omega _{0}$%
{\normalsize \ is usually in the microwave frequency region. For simplicity,
\ hereafter we will set }$\omega _{0}/\omega _{p}=10^{-5}${\normalsize . \
The other parameters are: \ }$\omega _{m}/\omega _{0}=4.0${\normalsize , \ }$%
\omega _{p}R/c=0.2${\normalsize , \ }$f_{2}=0.3${\normalsize , \ \ }$\alpha $%
{\normalsize \ is shown in the figures). \newline
\newline
Fig.2. (a) The frequency dependence of the effective permeability }$\mu
^{(-)}${\normalsize \ of magnetic grains and the corresponding frequency
dependencies of (b) the effective wave number }$k${\normalsize \ and (c) the
effective damping coefficient }$\beta ${\normalsize \ of the composite for
negative circularly polarized waves propagating in the direction of
magnetization. (The parameters are: }$\omega _{m}/\omega _{0}=4.0$%
{\normalsize , }$\omega _{p}R/c=0.2${\normalsize , \ }$f_{2}=0.3$%
{\normalsize , \ }$\alpha ${\normalsize \ is shown in the figures.) \newline
\newline
}

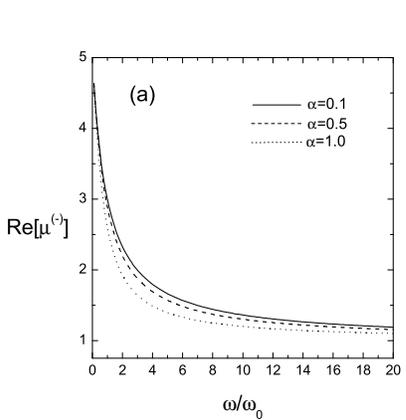
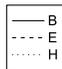
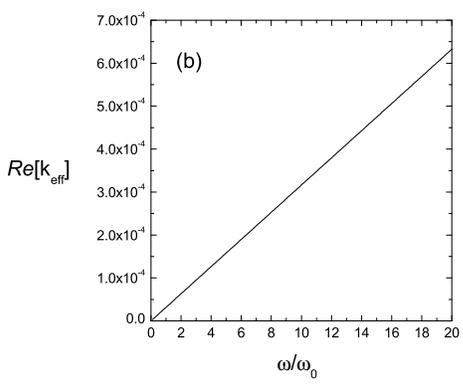
\end{document}